# Routing Light Emission from Monolayer MoS$_2$ by Mie Resonances of Crystalline Silicon Nanospheres

Keisuke Ozawa, Hiroshi Sugimoto*, Daisuke Shima, Tatsuki Hinamoto, Mojtaba Karimi Habil, Yan Joe Lee, Søren Raza, Keisuke Imaeda, Kosei Ueno, Mark L. Brongersma, and Minoru Fujii*

AUTHOR ADDRESS. Department of Electrical and Electronic Engineering, Graduate School of Engineering, Kobe University, Rokkodai, Nada, Kobe 657-8501, Japan

**ABSTRACT:** A dielectric Mie-resonant nanoantenna is capable of controlling the directionality of the emission from nearby quantum emitters through the excitation of multiple degenerate Mie resonances. A crystalline silicon nanosphere (Si NS) is a promising candidate for a dielectric nanoantenna because crystalline Si has a large refractive index (3.8 at 650 nm) and the small imaginary part of a complex refractive index (0.015 at 650 nm) as an optical material. In this work, we control the emission directionality of excitons supported by monolayer transition metal dichalcogenides (1L-TMDCs) using a Si NS. We first discuss the condition to extract the emission preferentially towards the Si NS side from the analytical calculations. We then study the photoluminescence (PL) of 1L-TMDCs on which differently sized single Si NSs are placed. We show that the PL spectral shape strongly depends on the emission direction, and that the emission toward the Si NS side (top) with respect to the opposite side (bottom) is the largest at wavelengths between the magnetic dipole and electric dipole Mie resonances of a Si NS. Finally, we quantitatively discuss the spectral shape of the top-to-bottom ratio from numerical simulations.

## INTRODUCTION

Atomically-thin transition metal dichalcogenides (TMDCs: e.g., MoS$_2$, MoSe$_2$, WS$_2$, WSe$_2$) display a wide range of desirable physical properties including stable and tunable excitonic behaviors, high electron mobility, and valley selectivity.[1-8] Among these phenomena, observation of room temperature excitonic luminescence in the visible to near-infrared range (1.5 - 2.5 eV)[9] in monolayer TMDCs (1L-TMDCs) has demonstrated their promise for use in solid state light emitting devices.[9-14] However, there remain several challenges such as a low light absorption, low emission quantum yield,[14,15] and low light-extraction efficiency that hinder the performance of light emitting devices. The usage of an optical nanoantenna is one of the possible routes to overcome these challenges. In fact, fluorescence enhancement of 1L-TMDCs has already been achieved by a plasmonic nanoantenna exhibiting localized surface plasmon resonances (LSPRs).[16-20] A limitation of a plasmonic nanoantenna is the lack of the capability to manipulate the emission directionality of a quantum emitter because of the almost solely electric dipole-type nature of the resonance unless complex structures such as a Yagi-Uda antenna are formed.[21-23]

In contrast to LSPRs of metallic nanoantennas, Mie resonances of high-index dielectric (e.g., silicon (Si) and gallium phosphide (GaP)) nanoantennas intrinsically support the magnetic-type as well as the electric-type multipolar modes.[24-30] Therefore, by properly designing the coupling between emitter dipoles and the multiple Mie resonances of a dielectric nanoantenna, the emission directionality can be controlled with a high degree of freedom.[26,31-33] Cihan et al.[31] succeeded in improving the light-extraction efficiency of 1L-TMDCs simply by placing a Si nanowire nanoantenna on top. Fang et al. used a hydrogenated amorphous Si nanosphere (a-Si:H NS) nanoantenna and demonstrated the capability of controlling the emission directionality.[32] A drawback of a-Si:H NS is the small refractive index (~2), which requires a relatively large size for emission directional control, and makes the local field enhancement and the Purcell factor small.[34]

In this work, we use a crystalline Si NS with the refractive index of ~3.8 as a small footprint dielectric nanoantenna to control the emission directionality of 1L-TMDCs. We first study the condition to maximize the emission directionality of a dipole transversely coupled to a Si NS from analytical calculations. We then experimentally investigate photoluminescence (PL) spectra of 1L-MoS$_2$ on which differently sized single Si NSs are placed. We detect the PL from the Si NS side (top) and the opposite side (bottom) and then obtain the top-to-bottom ratio of the PL spectra. We demonstrate that the largest top-to-bottom ratio is obtained at wavelengths between the magnetic dipole (MD) and electric dipole (ED) Mie resonances of a Si NS. Finally, we quantitatively analyze the top-to-bottom ratio spectra by comparing them with numerical simulations.

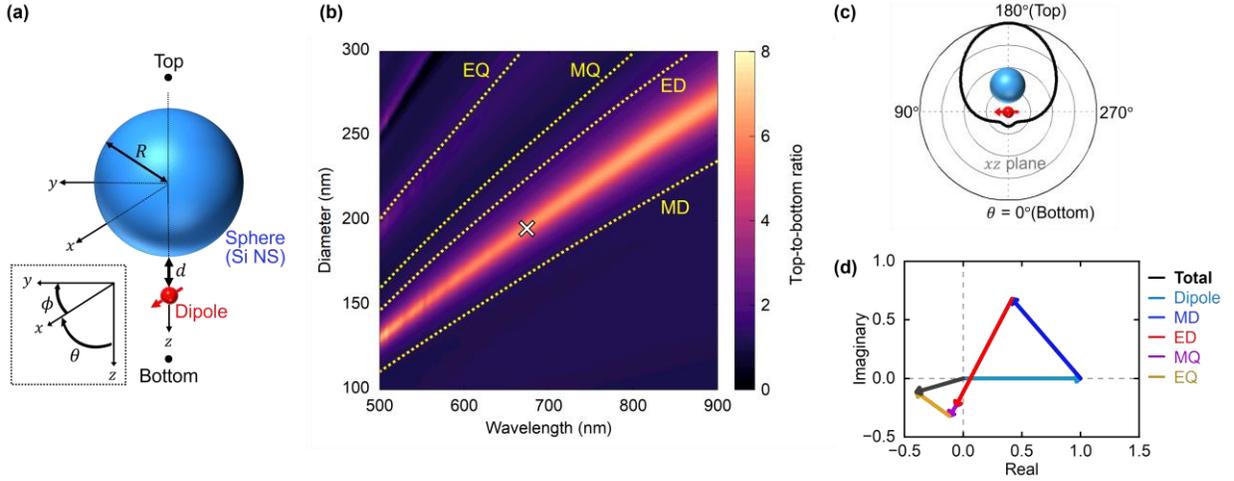

**Figure 1.** (a) Model structure for the calculation of the T/B ratio. A sphere of radius $R$ is located at the origin, and a dipole is positioned at the distance $d$ from the surface of the sphere on the $z$-axis. The dipole moment is oriented along the $x$-axis ($\mathbf{p}_{source} = p\mathbf{e}_x$). (b) Contour plot of the calculated T/B ratio as functions of wavelength and Si NS diameter. $d$ is set to 3 nm. Yellow dashed curves represent resonant wavelengths of the MD, ED, MQ and EQ resonances. A cross-mark is shown at 675 nm in wavelength and 196 nm in diameter. (c) Radiation pattern at 675 nm for a 196-nm-diameter Si NS. (d) Phasor diagram of the electric field in the bottom direction at 675 nm for a 196-nm-diameter Si NS. Labels indicate the contributions of each multipole, i.e., dipole, MD, ED, MQ and EQ resonances.

## RESULTS AND DISCUSSION

**Analytical calculation of the dipole radiation scattered by a Si NS.** Mie theory gives an analytical expression of the field scattered by a sphere under plane-wave illumination.[35,36] Similarly, the modified Mie theory gives the expression when a sphere is illuminated by an electric dipole source.[37,38] Figure 1a shows the model structure for the calculation of a scattered field under an electric dipole source illumination. A sphere of radius $R$ is located at the origin, and a dipole is positioned at the distance $d$ from the surface of the sphere on the $z$-axis. The dipole moment is oriented along the $x$-axis ($\mathbf{p}_{source} = p\mathbf{e}_x$). This mimics an in-plane dipole in 1L-TMDCs. The electric field is the sum of the dipole field ($\mathbf{E}^{Dip}$) and the scattered field ($\mathbf{E}^{Sca}$), and the intensity in the far-field ($I_{Tot}(\theta, \phi)$) is expressed as

$$I_{Tot}(\theta, \phi) \propto \left|E_\theta^{Dip} + E_\theta^{Sca}\right|^2 + \left|E_\phi^{Dip} + E_\phi^{Sca}\right|^2$$
$$\propto \left|\frac{k^2 p}{x_s}\sum_l \frac{2l+1}{l(l+1)}\{\cos\phi(\beta_l \pi_l + i\alpha_l \tau_l)\mathbf{e}_\theta - \sin\phi(\beta_l \tau_l + i\alpha_l \pi_l)\mathbf{e}_\phi\}\right|^2 \quad (1)$$

where $x_s = k(R + d)$ and $k$ is the vacuum wavenumber. $\pi_l = P_l^1/\sin\theta$ and $\tau_l = dP_l^1/d\theta$, where $P_l^1$ is the associated Legendre polynomial of degree $l$ and order 1. $\alpha_l$ and $\beta_l$ are modified Mie coefficients defined as $\alpha_l = \psi_l'(x_s) - a_l \xi_l'(x_s)$ and $\beta_l = \psi_l(x_s) - b_l \xi_l(x_s)$, where $a_l$ and $b_l$ are regular electric and magnetic Mie coefficients obtained using a plane-wave illumination, respectively. $\psi_l(z) = z j_l(z)$ and $\xi_l(z) = z h_l^1(z)$ are Riccati-Bessel functions derived from the spherical Bessel function $j_l$ and the Hankel function of the first kind $h_l^1$, respectively. Details on the derivation of eq 1 are described in Section 1 in Supporting Information.

To evaluate the emission directionality, we introduce the quantity "top-to-bottom ratio (T/B ratio)" defined as the intensity ratio between $\theta = \pi$ (top) and $\theta = 0$ (bottom) directions (both at $\phi = 0$),

$$\text{T/B ratio} = \frac{I_{Tot}(\theta=\pi,\phi=0)}{I_{Tot}(\theta=0,\phi=0)} = \frac{|\sum_n(-i)^n \frac{2n+1}{n(n+1)}\{\beta_n \pi_n(\pi) + i\alpha_n \tau_n(\pi)\}|^2}{|\sum_n(-i)^n \frac{2n+1}{n(n+1)}\{\beta_n \pi_n(0) + i\alpha_n \tau_n(0)\}|^2} \quad (2)$$

Figure 1b shows the calculated T/B ratio as functions of the wavelength and the Si NS diameter. The distance between the dipole and the surface of a Si NS is set to 3 nm. Yellow dashed curves represent resonant wavelengths of the MD, ED, magnetic quadrupole (MQ) and electric quadrupole (EQ) resonances of a Si NS. The T/B ratio is strongly enhanced between the MD and ED resonances and the maximum value reaches ~6. Figure 1c shows the radiation pattern of a 196-nm-diameter Si NS at 675 nm (cross-mark in Figure 1b). The bottom emission is strongly suppressed, and the emission mostly goes to the top direction. Figure 1d shows the phasor diagram of the electric field in the bottom direction. The incident field from the dipole is cancelled mainly by the MD and ED resonances. This results in the suppression of the total electric field in the bottom direction. On the other hand, the electric field intensity in the top direction is not strongly affected by the Si NS, although the phase is modified by the Mie resonances (see the phasor diagram in the top direction in Supporting information (Figure S1 in Section 2)).

**Experimental procedure and properties of 1L-MoS$_2$ and Si NSs.** We grew 1L-MoS$_2$ by chemical vapor deposition (CVD) and transferred it onto a silica (SiO$_2$) substrate for optical measurements.[39,40] Figure 2a,b shows the bright-field image and the PL image excited at 488 nm.

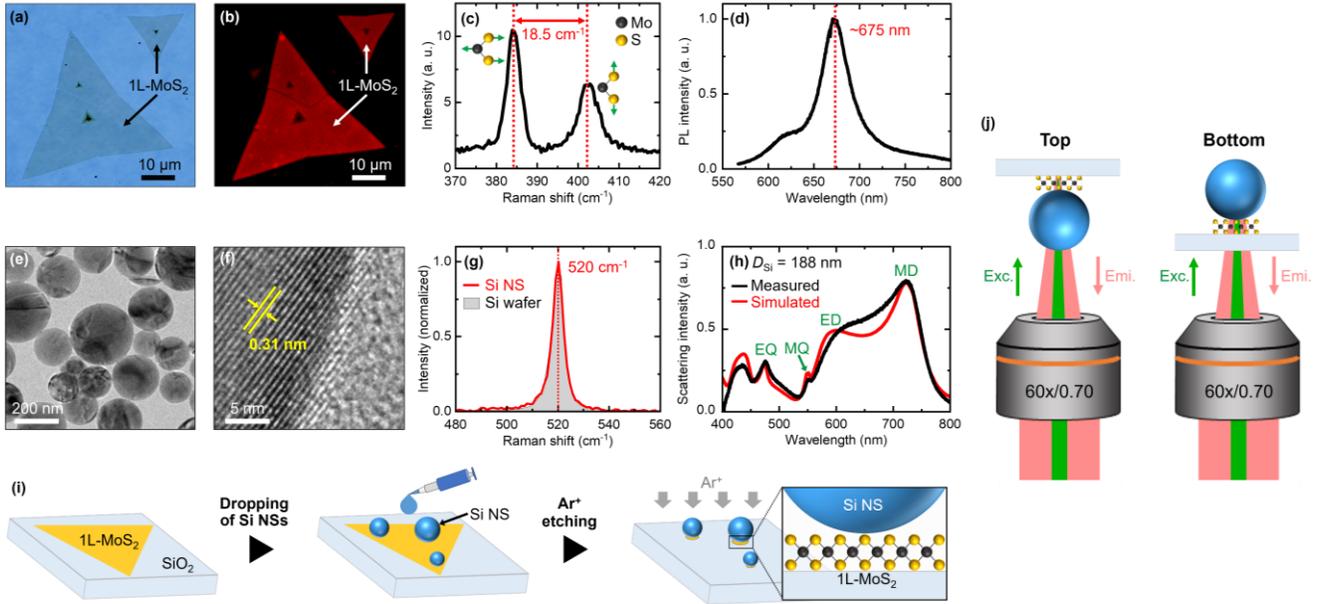

**Figure2.** (a) Bright-field image and (b) PL image excited at 488 nm of 1L-MoS$_2$ on a SiO$_2$ substrate. The scale bars are 10 μm. (c) Raman scattering spectrum of 1L-MoS$_2$ excited at 532 nm. The peak at 384.0 cm$^{-1}$ is the in-plan mode and that at 402.5 cm$^{-1}$ is the out-of-plan mode. (d) PL spectrum of 1L-MoS$_2$ excited at 532 nm. (e) TEM and (f) HRTEM images of Si NSs. The scale bars in (e) and (f) are 200 nm and 5 nm, respectively. (g) Raman scattering spectra of a single Si NS (red curve) and a Si wafer (grey area) excited at 532 nm. (h) Measured (black curve) and simulated (red curve) scattering spectra of a single Si NS with the diameter ($D_{Si}$) of 188 nm. (i) Schematic of the procedure to place Si NSs on 1L-MoS$_2$. (j) Schematic of the setup for PL measurements in the top and bottom directions.

Figure 2c shows the Raman scattering spectrum excited at 532 nm. The Raman peaks at 384.0 cm$^{-1}$ and 402.5 cm$^{-1}$ are assigned to the in-plane and out-of-plane modes, respectively. The separation of 18.5 cm$^{-1}$ between the two modes confirms the formation of a monolayer.[41,42] Figure 2d shows the PL spectrum excited at 532 nm. The exciton emission appears around 675 nm.

Figure 2e shows the transmission electron microscope (TEM) image of Si NSs used in this work. The preparation procedure is described in the METHOD section.[43] The Si NSs are almost perfectly spherical, and the diameters are distributed from 50 nm to 300 nm. Figure 2f shows the high resolution TEM (HRTEM) image. The lattice fringe corresponds to {111} planes of crystalline Si, demonstrating the high crystallinity. This can be confirmed by the Raman scattering spectrum in Figure 2g (red curve). The width of the Raman peak at 520 cm$^{-1}$ is as narrow as that of a Si wafer (grey area). Figure 2h shows a measured scattering spectrum of a single Si NS with the diameter ($D_{Si}$) of 188 nm (black curve). The MD, ED, MQ and EQ resonances are clearly observed. The measured spectrum agrees well with the simulated one (red curve), which guarantees the high quality of the Si NSs.

Figure 2i shows the procedure to place Si NSs on 1L-MoS$_2$. Diluted colloidal suspension of Si NSs is dropped onto 1L-MoS$_2$ formed on a SiO$_2$ substrate. The 1L-MoS$_2$ is then subjected to argon ion (Ar$^+$) etching using Si NSs as self-masks. This process leaves 1L-MoS$_2$ only below the Si NSs. The diameter of 1L-MoS$_2$ below a Si NS is slightly smaller than that of the Si NS as we will discuss later.

Figure 2j shows the setup for PL measurements in the top and bottom directions. To measure the PL toward the top direction, 1L-MoS$_2$ is excited from the Si NS side through an objective lens and the PL is collected from the same side by using the same objective lens. For the measurement of the PL toward the bottom direction, the same setup is used except that the sample is flipped, i.e., 1L-MoS$_2$ is excited from a SiO$_2$ substrate side and the PL is collected from the same side.

**Directional emission of 1L-MoS$_2$ with a Si NS nanoantenna.** Figure 3a shows the scattering spectrum of a Si NS ($D_{Si}$ = 194 nm) on 1L-MoS$_2$ together with the PL spectrum of a pristine 1L-MoS$_2$ (grey area). Figure 3b shows the PL spectra of 1L-MoS$_2$ with a Si NS ($D_{Si}$ = 194 nm) measured in the top (red curve) and bottom (blue curve) directions. In both cases, the spectral shape is strongly modified from that of a pristine 1L-MoS$_2$. For example, an additional peak appears around 755 nm both in the top and bottom directions. Since the wavelength coincides with the MD resonance in the scattering spectrum, it is due to the Purcell effect induced by the MD resonance. In addition, the spectral shape in the top and bottom directions is significantly different in the 600 – 700 nm range. Figure 3c shows the T/B ratio spectrum obtained by dividing the PL spectrum in the top direction by that in the bottom direction. A clear peak appears in the T/B ratio spectrum and the maximum value is ~3 around 645 nm. Since the T/B ratio is 0.58 without a Si NS (magenta dashed line), it is enhanced ~5.2-fold in the presence of a Si NS. Figure 3d-f shows the same dataset for a Si NS ($D_{Si}$ = 213 nm). In Figure 3e, the Purcell effect induced by the MQ resonance is clearly observed around 605 nm. Moreover, the spectral shape is different between the two directions in the 620 – 750 nm range. In Figure 3f, the peak of the T/B ratio spectrum appears around 695 nm and the maximum value reaches ~7, giving an enhancement factor of ~12.

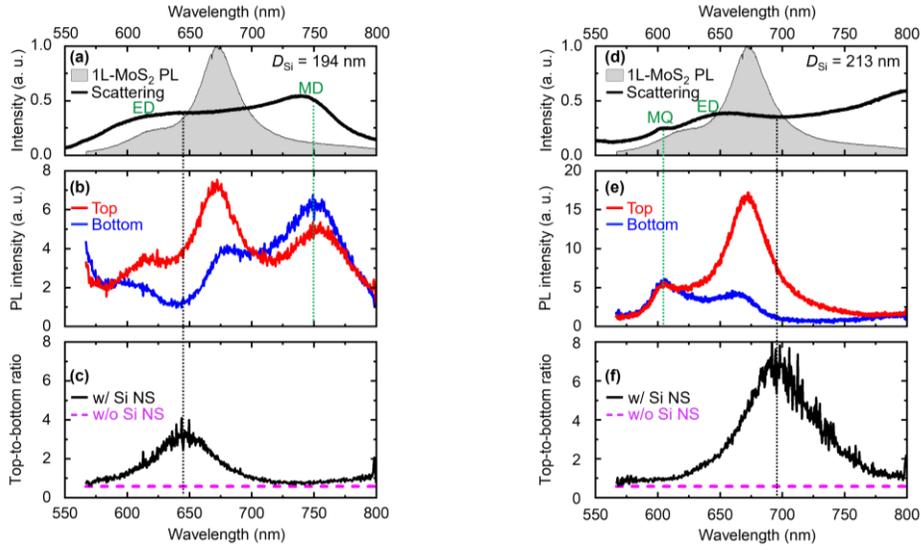

**Figure 3.** (a) Scattering spectrum of a Si NS ($D_{Si}$ = 194 nm) on 1L-MoS$_2$ (black curve) together with the PL spectrum of pristine 1L-MoS$_2$ (grey area). (b) PL spectra of 1L-MoS$_2$ with a Si NS ($D_{Si}$ = 194 nm) measured in the top (red curve) and bottom (blue curve) directions. (c) T/B ratio spectra of 1L-MoS$_2$ with (black curves) and without (Magenta dashed lines) a Si NS. (d-f) The same dataset as (a-c) for a 213-nm-dimeter Si NS.

We perform the same measurements for different size Si NSs. Figure 4a shows the T/B ratio spectra (red dots) and the scattering spectra (grey shaded curves) of 1L-MoS$_2$ with single Si NSs. The diameter of the Si NSs is changed from 166 nm to 248 nm. The T/B ratio peak appears always between the MD and ED resonances. In Figure 4b, the peak wavelength of the T/B ratio spectrum (black dots) is plotted as a function of the Si NS diameter. The peak shifts to the longer wavelength with increasing the diameter. The blue line in Figure 4b shows the T/B ratio peak calculated using eq 2. We can see good agreement between the measured and calculated results.

In Figure 4a, when the diameter of a Si NS is 248 nm, the T/B ratio spectrum has an additional peak near the EQ resonance (around 600 nm). The radiation pattern and the phasor diagrams of the electric field in the top and bottom directions at the peak wavelength are shown in Supporting information (Figure S2 in Section 3). The data indicate that the peak arises from the interference with the higher-order Mie resonances, mainly the EQ resonance.

In Figure 4a, we notice that the maximum value of the T/B ratio depends on the size of a Si NS, while it takes almost the same value (~6) in analytical calculations in Figure 1b. In order to quantitatively analyze the observed T/B ratio spectra, we carry out numerical simulation, where we consider the emission and excitation processes separately. For the simulation of the emission process, we consider the model structure in Figure 5a. A single in-plane dipole source is placed 1 nm above a SiO$_2$ substrate, and a Si NS is placed 3 nm above the dipole source. We calculate the emitted power collected within N.A. = 0.7 in the top and bottom directions and take the ratio (emission T/B ratio). For the simulation of the excitation process in Figure 5b, we launch a plane-wave (532 nm in wavelength) from the top and bottom directions, calculate the in-plane component of the electric field intensity ($|\mathbf{E}_\parallel|^2/|\mathbf{E_0}|^2$) under a Si NS, and take the ratio (excitation T/B ratio).

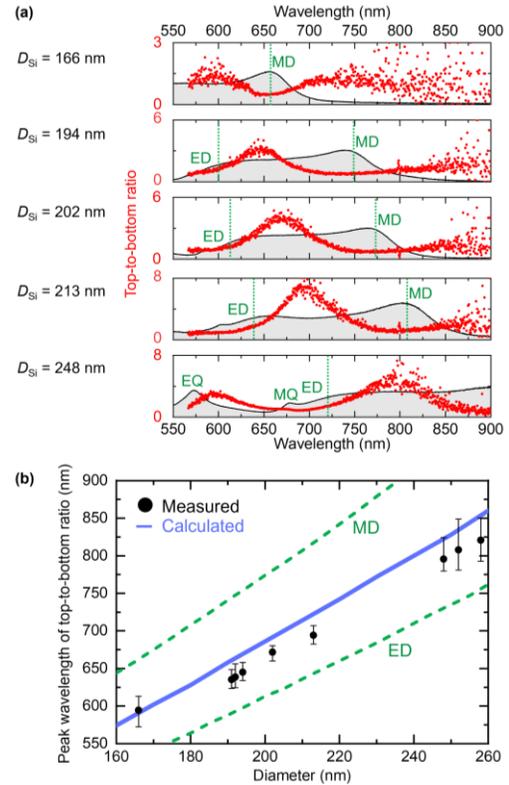

**Figure 4.** (a) T/B ratio (red dots) and scattering (grey areas) spectra of 1L-MoS$_2$ with differently sized single Si NSs. Green dashed lines represent the resonant wavelengths of the MD and ED resonances. (b) Peak wavelengths of the T/B ratio spectra as a function of the diameter of a Si NS (black dots). Blue line shows the T/B ratio peak wavelength calculated using eq (2). Green dashed lines represent the resonant wavelengths of the MD and ED resonances.

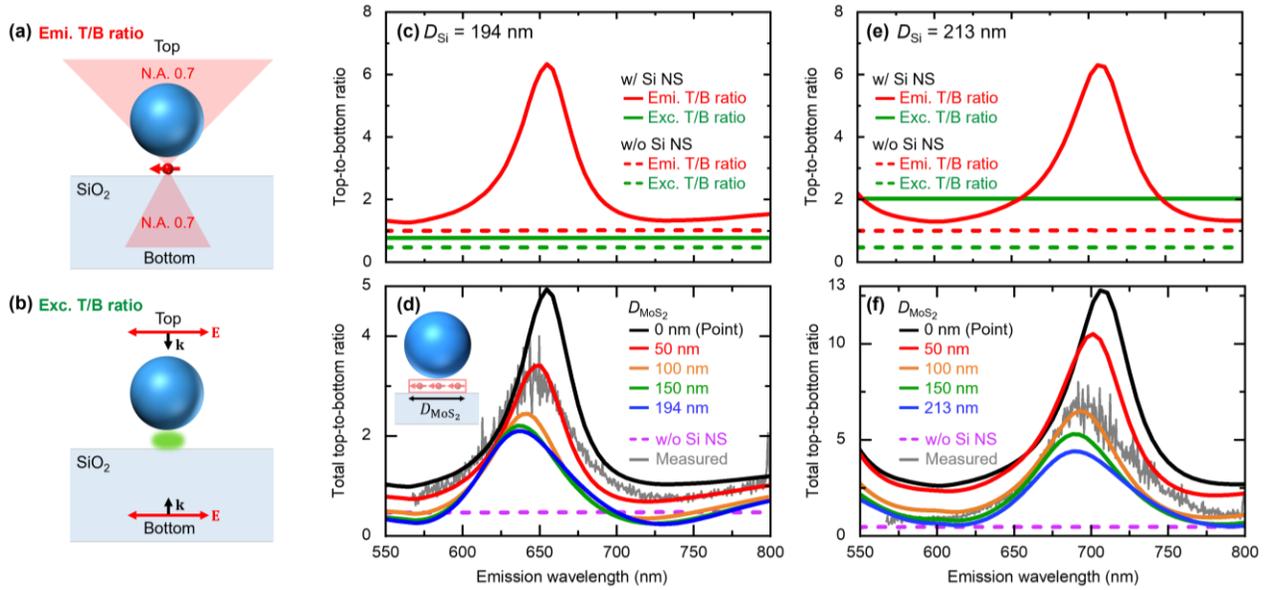

**Figure 5.** (a,b) Schematic of the simulation setups for the calculation of the (a) emission (Emi.) and (b) excitation (Exc.) T/B ratios. (c) Emission (red solid curve) and excitation (green solid line) T/B ratios for $D_{Si}$ = 194 nm. (d) Total T/B ratio for $D_{Si}$ = 194 nm for different values of $D_{MoS_2}$. Inset shows the definition of $D_{MoS_2}$. Magenta dashed curve and grey solid curve show the result obtained without a Si NS and measured T/B ratio, respectively. (e,f) The same dataset as (c,d) for a 213-nm-dimeter Si NS.

Figure 5c shows the emission T/B ratio spectrum for a 194 nm-diameter Si NS (red solid curve) together with that without a Si NS (red dashed curve). The emission T/B ratio has a peak at 655 nm and the maximum value is ~6.3. In the same figure, the excitation T/B ratios at 532 nm with and without a Si NS are shown by horizontal green lines. The excitation T/B ratio is slightly enhanced in the presence of a Si NS. In Figure 5d (black curve), the total T/B ratio spectrum obtained by multiplying the excitation and emission T/B ratios is shown. The maximum value is ~5 around 655 nm. This value is larger than the experimentally obtained one (~3 in Figure 3c), and the peak wavelength is slightly longer than the experimental maximum wavelength (~645 nm). We will discuss the origin of the discrepancy later.

Figure 5e shows the data obtained for a 213-nm-diameter Si NS. In this case, the excitation T/B ratio at 532 nm is ~2. This value is similar to the emission T/B ratio at 532 nm for a 213-nm-diameter Si NS in Figure 1b (~1.8). This is not a coincidence, but is explained by the reciprocity theorem.[44] Figure 5f (black curve) shows the total T/B ratio spectrum of a 213-nm-diameter Si NS. The total T/B ratio reaches ~13 at 705 nm. This value is again larger than the experimentally obtained one (~7 in Figure 3f), and the peak wavelength is longer than the experimental peak wavelength (~695 nm).

A possible origin of the discrepancy in the T/B ratio between the experiments and the simulations is the unrealistic assumption that a single point dipole exists below a Si NS. We therefore consider a model where dipoles are distributed in a finite area, i.e., a circle with the diameter $D_{MoS_2}$, below a Si NS as schematically shown in the inset in Figure 5d (see Section 4 in Supporting Information for details). In Figure 5d and Figure 5f, the total T/B ratio spectra obtained for different values of $D_{MoS_2}$ are shown. The total T/B ratio becomes smaller and the peak shifts to the shorter wavelength with increasing $D_{MoS_2}$. This arises from the dipole position dependence of the emission spectra (see Figure S4 in Supporting Information). In Figure 5d and 5f, the experimental T/B ratio spectra agree well with the calculated ones when $D_{MoS_2}$ is 50 nm and 100 nm, respectively. These values are smaller than the diameter of the Si NSs due to side-etching during the etching process.

## CONCLUSIONS

We demonstrated that a crystalline Si NS with low-order Mie resonances at the wavelength of 600 - 700 nm can serve as a small-footprint directional nanoantenna to enhance the top emission of 1L-TMDCs. Analytical calculations revealed that the directional emission arises from the interference between an emitting dipole and multiple Mie resonances. We measured PL spectra of 1L-MoS$_2$ on which different size single Si NSs are placed, and observed the significant difference in the top and bottom directions. The T/B ratio reached ~7. The observed size dependence of the T/B ratio spectra could be well reproduced by numerical simulations.

## METHODS

**Preparation for the colloidal suspension of Si NSs.** Si NSs were prepared using a previously reported method that utilizes thermal disproportionation of silicon monoxide (SiO).[43] SiO lumps (several mm in size) (99%, FUJIFILM Wako) were annealed at 1500 °C in a $N_2$ gas atmosphere for 30 min to grow crystalline Si NSs in $SiO_2$ matrices. Si NSs were liberated by dissolving $SiO_2$ matrices in hydrofluoric acid (HF) solution (46 wt%) for 1 h. After liberating Si NSs, HF was removed by several centrifugation processes. Finally, Si NSs were transferred to methanol and subjected to ultrasonication for 1 min with an ultrasonic homogenizer (Violamo SONICSTAR 85)].

**Procedure to place Si NSs on 1L-MoS$_2$.** A diluted colloidal suspension of Si NSs (0.25 mg/ml in methanol) was dropped onto 1L-MoS$_2$ on a $SiO_2$ substrate and dried in air. The substrate was then subjected to argon ion etching (ANELVA, L-201D) by using Si NSs as self-masks. This process leaves 1L-MoS$_2$ only below Si NSs.

**Scattering and PL spectroscopy of single Si NSs and 1L-MoS$_2$ with Si NSs.** An inverted optical microscope was used to measure the scattering spectra of the single Si NSs. A $SiO_2$ substrate with Si NSs and 1L-MoS$_2$ was placed face down onto the stage and illuminated with a halogen lamp through a dark-field objective lens (100x, N.A. = 0.9). Scattered light was collected by the same objective lens and transferred to the entrance slit of a spectrometer (Kymera 328i, Andor) and detected using a cooled CCD (Newton, Andor). To measure the PL spectra in the top direction, a substrate was placed face down onto the stage of a microscope and illuminated with 532 nm laser (Sapphire SF NX532, Coherent) through an objective lens (60x, N.A. = 0.7). The emitted light was collected by the same objective lens and transferred to the entrance slit of a spectrometer (SpectraPro-300i, Acton Research Corp.) and detected using a liquid-$N_2$-cooled CCD (Princeton Instruments). The PL spectra in the bottom direction were measured by the same setup by flipping the substrate upside down, i.e., the substrate was placed face up onto the stage.

**Simulation for scattering spectrum of single Si NSs.** A scattering spectrum of the single Si NSs on a $SiO_2$ substrate was calculated by the finite element method (FEM) using COMSOL Multiphysics. A Si NS was illuminated with a plane wave at the incident angle of 75° to the normal to a substrate, and the scattered time-averaged power flow over a cap of 64°, corresponding to the numerical aperture of the objective lens (N.A. = 0.9), was calculated. To avoid artificial reflections from the outer boundaries, we truncated the model using spherical perfectly matched layer (PML) with a thickness of half-wavelength. We applied a free tetrahedral mesh throughout the simulation domains, with the maximum element mesh size of $\lambda/10$ in air and in the substrate, and $D_{Si}/8$ in the Si NS domain.

**Simulation of T/B ratios in the emission and excitation processes.** A commercial FDTD simulation software (Lumerical, Ansys) was used for the simulation of the T/B ratio. In the emission process, a single in-plane dipole source was placed 1 nm above a $SiO_2$ substrate and a Si NS was placed 3 nm above the dipole source. Power emitted from the dipole was detected utilizing 2D monitors placed 3 μm from the center of a Si NS in the top and bottom directions. The Poynting vector was integrated over the detection angles corresponding to N.A. = 0.7 after a near- to far-field transformation of the captured field. To consider the finite area of 1L-MoS$_2$, a dipole was moved along the x-axis at 2 nm intervals to $D_{MoS_2}/2$. The power emitted from dipoles oriented in x- and y-directions was integrated within the circle. In the excitation process, a plane-wave was launched from the top or bottom directions, and in-plane components of the electric field intensity ($|\mathbf{E}_\parallel|^2/|\mathbf{E_0}|^2$) were calculated at 1 nm above a $SiO_2$ substrate. Considering the finite area of 1L-MoS$_2$, $|\mathbf{E}_\parallel|^2/|\mathbf{E_0}|^2$ was averaged within a circle with the diameter of $D_{MoS_2}$. The refractive index of Si was taken from the literature[45] and that of $SiO_2$ was set to 1.46.

## ASSOCIATED CONTENT

**Supporting Information.** This Supporting information is available free of charge at http://pubs.acs.org.
Additional data and discussion on the analytical calculation of the electric field and emission intensity under irradiation of a sphere by an electric dipole source and on the simulated results of the emission and excitation T/B ratios.

## AUTHOR INFORMATION

### Corresponding Author

**Hiroshi Sugimoto** – *Department of Electrical and Electronic Engineering, Graduate School of Engineering, Kobe University, 1-1 Rokkodai, Nada, Kobe 657-8501, Japan; orcid.org/0000-0002-1520-0940; Email: sugimoto@eedept.kobe-u.ac.jp*

**Minoru Fujii** – *Department of Electrical and Electronic Engineering, Graduate School of Engineering, Kobe University, 1-1 Rokkodai, Nada, Kobe 657-8501, Japan; orcid.org/0000-0003-4869-7399 Email: fujii@eedept.kobe-u.ac.jp*

### Authors

**Keisuke Ozawa** – *Department of Electrical and Electronic Engineering, Graduate School of Engineering, Kobe University, 1-1 Rokkodai, Nada, Kobe 657-8501, Japan*
**Daisuke Shima** – *Department of Electrical and Electronic Engineering, Graduate School of Engineering, Kobe University, 1-1 Rokkodai, Nada, Kobe 657-8501, Japan*
**Tatsuki Hinamoto** – *Department of Electrical and Electronic Engineering, Graduate School of Engineering, Kobe University, 1-1 Rokkodai, Nada, Kobe 657-8501, Japan; orcid.org/0000-0001-6607-2574*
**Mojtaba Karimi Habil** – *Department of Electrical and Electronic Engineering, Graduate School of Engineering, Kobe University, 1-1 Rokkodai, Nada, Kobe 657-8501, Japan*
**Yan Joe Lee** – *Geballe Laboratory for Advanced Materials, Stanford University, Stanford, California 94305, United States*
**Søren Raza** – *Department of Physics, Technical University of Denmark, DK-2800 Kongens Lyngby, Denmark*
**Keisuke Imaeda** – *Department of Chemistry, Faculty of Science, Hokkaido University, Kita 10, Nishi 8, Kita-ku, Sapporo, Hokkaido 060-0810, Japan*
**Kosei Ueno** – *Department of Chemistry, Faculty of Science, Hokkaido University, Kita 10, Nishi 8, Kita-ku, Sapporo, Hokkaido 060-0810, Japan*

**Mark L. Brongersma** - *Geballe Laboratory for Advanced Materials, Stanford University, Stanford, California 94305, United States; orcid.org/0000-0003-1777-8970*

Notes

The authors declare no competing financial interest.

ACKNOWLEDGMENT

This work is partially supported by JSPS KAKENHI Grant Numbers 21H01748, 22K18949 and 24K01287 and Kobe University Strategic International Collaborative Research Grant (Type B and C). Y. J. Lee and M .L. B. acknowledge support from the Department of Energy Grant DE-FG07-ER46426.